\documentclass[english, aps, pre, twocolumn]{revtex4-1}
\usepackage[T1]{fontenc}
\usepackage[latin9]{inputenc}
\usepackage{color}
\usepackage{babel}
\usepackage{float}
\usepackage{amsmath}
\usepackage{amssymb}
\usepackage{graphicx}
\usepackage[unicode=true,pdfusetitle,
 bookmarks=true,bookmarksnumbered=false,bookmarksopen=false,
 breaklinks=false,pdfborder={0 0 0},pdfborderstyle={},backref=false,colorlinks=true]{hyperref}
\hypersetup{linkcolor=darkblue, citecolor=darkblue, urlcolor=darkblue}
\usepackage{breakurl}

\makeatletter
\renewcommand{\fnum@figure}{FIG.~\thefigure}

\definecolor{darkblue}{rgb}{0.0, 0.0, 0.55}

\@ifundefined{showcaptionsetup}{}{
 \PassOptionsToPackage{caption=false}{subfig}}
\usepackage[caption=false]{subfig}
\makeatother

\begin{document}

\title{Radiative transfer dynamo effect}

\author{Vadim R. Munirov}
\email{vmunirov@pppl.gov}
\affiliation{Princeton Plasma Physics Laboratory, Princeton University, Princeton,
New Jersey 08543, USA}
\affiliation{Department of Astrophysical Sciences, Princeton University, Princeton,
New Jersey 08540, USA}

\author{Nathaniel J. Fisch}
\affiliation{Princeton Plasma Physics Laboratory, Princeton University, Princeton,
New Jersey 08543, USA}
\affiliation{Department of Astrophysical Sciences, Princeton University, Princeton,
New Jersey 08540, USA}

\date{Received 6 September 2016; published 17 January 2017}

\begin{abstract}
Magnetic fields in rotating and radiating astrophysical plasma can
be produced due to a radiative interaction between plasma layers moving
relative to each other. The efficiency of current drive, and with
it the associated dynamo effect, is considered in a number of limits.
It is shown here, however, that predictions for these generated magnetic
fields can be significantly higher when kinetic effects, previously
neglected, are taken into account.
\end{abstract}

\maketitle

\section{Introduction}

Cosmic magnetism is usually explained by magnetohydrodynamic dynamo theory \cite{Brandenburg2005},
which is, however, only an amplification mechanism that still requires
some initial seed field. There have been many speculations about the
origin of the seed field, but consensus is still lacking \cite{Widrow2002,Raychaudhuri1972}.
One possible mechanism is a radiation induced drag force on electrons
in rotating astrophysical objects. This idea was evidently first proposed
by Cattani and Sacchi \cite{Cattani1966} and later has been applied
to different astrophysical conditions and objects \cite{Harrison1970,MishustinRuzmaikin1972,HinataDaneshvar1984,Walker1988,Contopoulos1998,Contopoulos2006,Contopoulos2015,Bisnovatyi-Kogan1977,Bisnovatyi-Kogan2002,AndoDoiSusa2010,Langer2003,Durrive2015,Gopal2005}.
However, none of these studies took into account kinetic effects.
It is shown here that predictions for these generated magnetic fields
can be significantly higher when kinetic effects are taken into account.
In the presence of existing magnetic fields, these kinetic
effects can enhance the generated magnetic fields by orders of magnitude.

A rotating astrophysical object is subject to asymmetric incoming
radiation, which exerts the Poynting-Robertson drag force on electrons
in the (toroidal) rotation direction that leads to the poloidal magnetic
field. Within a fluid framework, this can be modeled by including
an additional term into the equation for the magnetic field dynamics:

\begin{equation}
\frac{\partial\mathbf{B}}{\partial t}=-\frac{c}{e}\nabla\times\mathbf{f}_{rad}.
\end{equation}

There are two ways in which kinetic effects modify the effective radiation
force. First, the Poynting-Robertson force on an individual electron
depends on the absorbed power, which is, generally speaking, different
for the electrons of different energies; usually the more energetic electrons
absorb more power. Thus, to get the effective radiation force on the
electron fluid, one needs to average the force for each electron over
the absorbed power. Second, toroidal current can be driven even without
toroidal momentum injection just by asymmetrically heating electrons.
Indeed, by heating electrons we increase their energy and since the
collision frequency in plasma is energy dependent ($\propto v^{-3}$)
the toroidal drag force due to Coulomb collisions is going to be asymmetric
resulting in the total toroidal current \cite{FischBoozer1980}.

\begin{figure}[]
\includegraphics[scale=0.33]{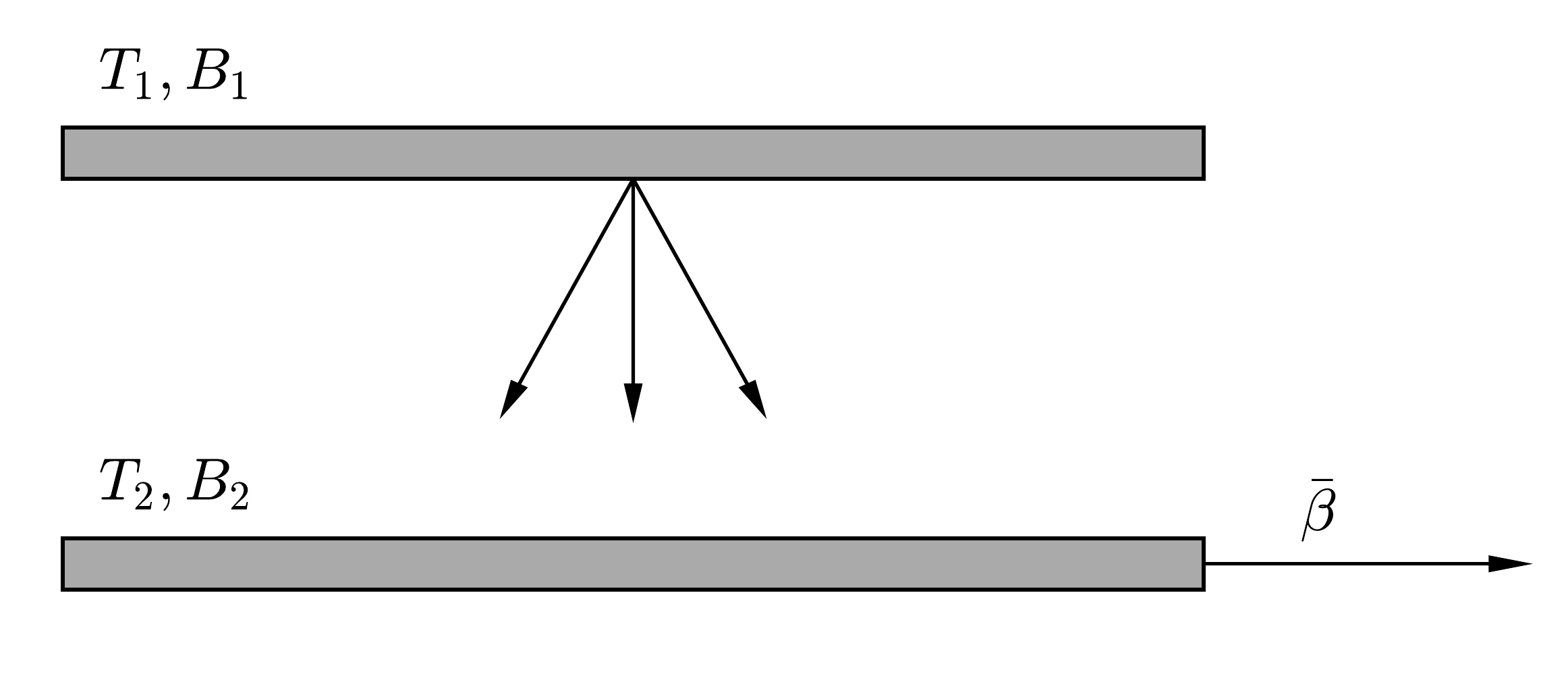}
\setlength{\belowcaptionskip}{-10pt}
\caption{Parallel radiating and absorbing slabs of plasma, immersed in different magnetic fields at different temperatures, in relative parallel motion.
\label{fig01}}
\end{figure}

To simplify the problem and underscore the influence of the kinetic
effects, we consider a slab geometry, where the parallel direction
corresponds to the toroidal direction of the original rotating object
(see Fig.~\ref{fig01}). Namely, we consider two parallel and possibly
magnetized (in the parallel direction) plasma slabs that move relative
to each other with velocity $\bar{\beta}$ (velocities are measured
in the units of $c$). We label the upper slab slab 1 and the lower
slab slab 2.

\setlength{\belowcaptionskip}{0pt}

The paper is organized as follows:  In Sec.~II we derive the efficiency of current generation through the Poynting-Robertson effect.
In Sec.~III,  using kinetic rather than fluid theory, we show how the efficiency of the current generation through radiation effects can be much enhanced when there is a seed magnetic field already present and when kinetic effects are considered.  We consider, in Sec.~IIIA, the case of blackbody emission and cyclotron absorption. In Sec.~IIIB, we consider the case of cyclotron emission and cyclotron absorption, where not only can the currents driven be driven much more effectively, but there is even the curious effect that the current in adjacent differentially moving plasma can be either in the same direction or in opposite directions.  In Sec.~IV, we summarize and discuss our findings. 

\section{The Poynting-Robertson effect}

Consider an electron that moves with velocity $\beta_{\Vert}$ and
emits isotropic radiation in its own reference frame. Imagine that
this electron also absorbs external radiation, which is isotropic
in its own reference frame moving with parallel velocity $\beta_{s}=-\bar{\beta}$.
Conservation of energy and momentum then gives

\noindent 
\begin{align}
 & mc\left(\gamma\dot{\beta}_{\parallel}+\dot{\gamma}\beta_{\parallel}\right)+\dot{p}_{\parallel}^{ems}=\dot{p}_{\parallel}^{abs},\\
 & mc^{2}\dot{\gamma}=P^{abs}-P^{ems},
\end{align}

\noindent where $P^{abs}$ is the absorbed power, $p_{\parallel}^{abs}$
is the absorbed parallel momentum, $P^{ems}$ is the emitted power,
and $p_{\parallel}^{ems}$ is the emitted parallel momentum.

The time derivative of the wave momentum is determined by the power
delivered by the wave $\dot{\mathbf{p}}^{wave}=\left(\mathbf{k}/\omega\right)P^{wave}$.
Using the Lorentz transformation we can express it as $\dot{p}^{ems}=\left(\beta_{\parallel}/c\right)P^{ems},$
$\dot{p}^{abs}=\left(\beta_{s}/c\right)P^{abs}.$ Inserting these
expressions into the energy-momentum equations we find that electron
parallel velocity satisfies 
\begin{equation}
\dot{\beta}_{\parallel}=-\frac{P^{abs}}{\gamma mc^{2}}(\beta_{\parallel}-\beta_{s}).
\end{equation}

We see that the electron experience drag by absorbing the external
radiation. This effect is called the Poynting\textendash Robertson
effect. It is a relativistic effect by its very nature, although it does
not require that the relative velocity between absorber and emitter  be relativistic.
In the reference frame of an electron, this drag force can be simply
interpreted as a momentum transfer from asymmetric external radiation
to an electron. However, in the reference frame of the external radiation
source, there is no parallel momentum injection; there is only an
energy increase, which makes the electron heavier relativistically.
Since the total parallel momentum must be conserved, the electron must
slow down. Notice that, without external radiation, there would be
no radiation drag force (if we define force as the cause of velocity
change rather than momentum), which is consistent with the well-known
fact that isotropically radiating charge conserves its parallel velocity
\cite{Zheleznyakov1996}. It should be emphasized that here, by absorption,
we mean a generalized process of wave-particle interaction; for example,
it can denote  Thomson scattering. While electrons experience radiation
drag, ions are almost unaffected by radiation and hence current is
generated.

Let us estimate crudely the current drive efficiency in this case. 
Assume that the parallel velocity is randomized during
the inverse collision time $\nu^{-1}$ and use the effective electron-electron
and electron-proton Coulomb collision frequency $\nu=6\Gamma/\beta^{3}$
(see Refs. \cite{Bornatici1995,FidoneGranataJohner1988,Fisch1981}),
where\textbf{ $\Gamma=\omega_{p}^{4}\ln\varLambda/4\pi nc^{3}$}.
Then, after averaging over the Maxwell distribution, we find:

\begin{equation}
\frac{j_{\parallel}}{P_{V}^{abs}}=\frac{\left\langle \beta^{3}\right\rangle }{15}\bar{\beta}\approx0.43\beta_{th}^{3}\bar{\beta}.\label{eq:eff_PR}
\end{equation}

\noindent Here and later the current drive efficiency is expressed
in the units of $e/\Gamma mc$ except for Eq.~(\ref{eq:eff_incremental}).

\section{Kinetic formulation}

The kinetic theory of current drive by external radiation is well
developed and experimentally demonstrated \cite{Fisch1987}. This
theory has been advanced to accommodate the need for efficient non-inductive
toroidal current generation required for the successful operation
of commercial tokamaks. The theory formulates the efficiency of current
generation as the ratio of the driven current density to the absorbed
power density \cite{Fisch1981}:

\begin{equation}
\frac{j_{\parallel}}{P_{V}^{abs}}=-\frac{e}{mc}\left[\frac{n_{\parallel}}{\nu}+\frac{\beta_{\parallel}}{\beta}\frac{\partial}{\partial\beta}\left(\frac{1}{\nu}\right)\right].\label{eq:eff_incremental}
\end{equation}

\noindent The first term in Eq.~(\ref{eq:eff_incremental}) can be
associated with the Poynting-Robertson drag, while the second is due
to asymmetric heating. The second term arises because the collision
frequency $\nu$ depends sensitively on the electron energy. It leads
to the electron cyclotron current drive effect in tokamaks
\cite{FischBoozer1980}. The radiative transfer dynamo effect in astrophysical
contexts is not much different from the situation described above.
The major difference is that the radiation driving current is set
up naturally rather than controlled.

Equation (\ref{eq:eff_incremental}) gives the non-relativistic efficiency
of the current driven by a very narrow radiation band that affects
only a small region in velocity space. To calculate the efficiency
for arbitrary incoming radiation average Eq.~(\ref{eq:eff_incremental})
over the power density absorbed per frequency per solid angle per
$d^{3}\boldsymbol{\beta}$. The absorbed power density is given by

\begin{equation}
P_{V}^{abs}=\iint d\omega d\Omega\alpha_{\omega\Omega}I_{\omega\Omega},\label{eq:P_abs}
\end{equation}

\noindent where $I_{\omega\Omega}$ is the incoming electromagnetic
energy flux density per unit frequency per solid angle and $\alpha_{\omega\Omega}$
is the absorption coefficient (true absorption plus stimulated emission).
Due to the principle of detailed balance the absorption coefficient
can be expressed through the emissivity of an individual electron
$\eta_{\omega\Omega}\left(\mathbf{p}\right)$ \cite{Trubnikov1963,BornaticiCanoDeBarbieriEngelman1983}:

\begin{equation}
\alpha_{\omega\Omega}=-\frac{8\pi^{3}c^{2}}{n_{r}^{2}\omega^{2}}\int d^{3}\mathbf{p}\eta_{\omega\Omega}\left(\mathbf{p}\right)\left(\frac{\partial f}{\partial\varepsilon}+\frac{n_{\parallel}}{c}\frac{\partial f}{\partial p_{\parallel}}\right),\label{eq:coef_abs}
\end{equation}

\noindent where $n_{r}$ is the ray refractive index, we will use
approximation of tenuous plasma and assume $n_{r}\approx1$; $n_{\parallel}$
is the wave parallel refractive index, which we take to be just $n_{\parallel}=\cos\theta$.

Thus, we can calculate the current drive efficiency for a specific
type of the absorption mechanism determined by $\eta_{\omega\Omega}\left(\mathbf{p}\right)$
and for a given external radiation spectrum $I_{\omega\Omega}$.
Although, in both emitting and absorbing radiation, the two slabs
form a coupled system, to get the efficiency linear in power transferred,
note that each slab may be considered to see fixed radiation from the
other slab.

We first argue that it is the current drive efficiency that determines large-scale
magnetic field generation in optically thick plasma, for which the
effect is maximized. For optically thick plasma, the incoming radiation
flux $I=\iint d\omega d\Omega I_{\omega\Omega}$ is absorbed over
the characteristic distance $R=\alpha^{-1}$, where $\alpha=P_{V}^{abs}/I$
is the characteristic absorption coefficient. Ampere's law gives $B\cdot2\pi r\approx\left(4\pi/c\right)j_{\parallel}Rh$,
where $h$ is the height of the plasma disk, so the large-scale equilibrium
magnetic field at distance $r$ outside the plasma is proportional
to the current drive efficiency:

\begin{equation}
B\approx\frac{2}{c}\frac{h}{r}\left(\frac{j_{\parallel}}{P_{V}^{abs}}\right)I.
\end{equation}

\begin{figure}[]
\includegraphics[width=8.6cm]{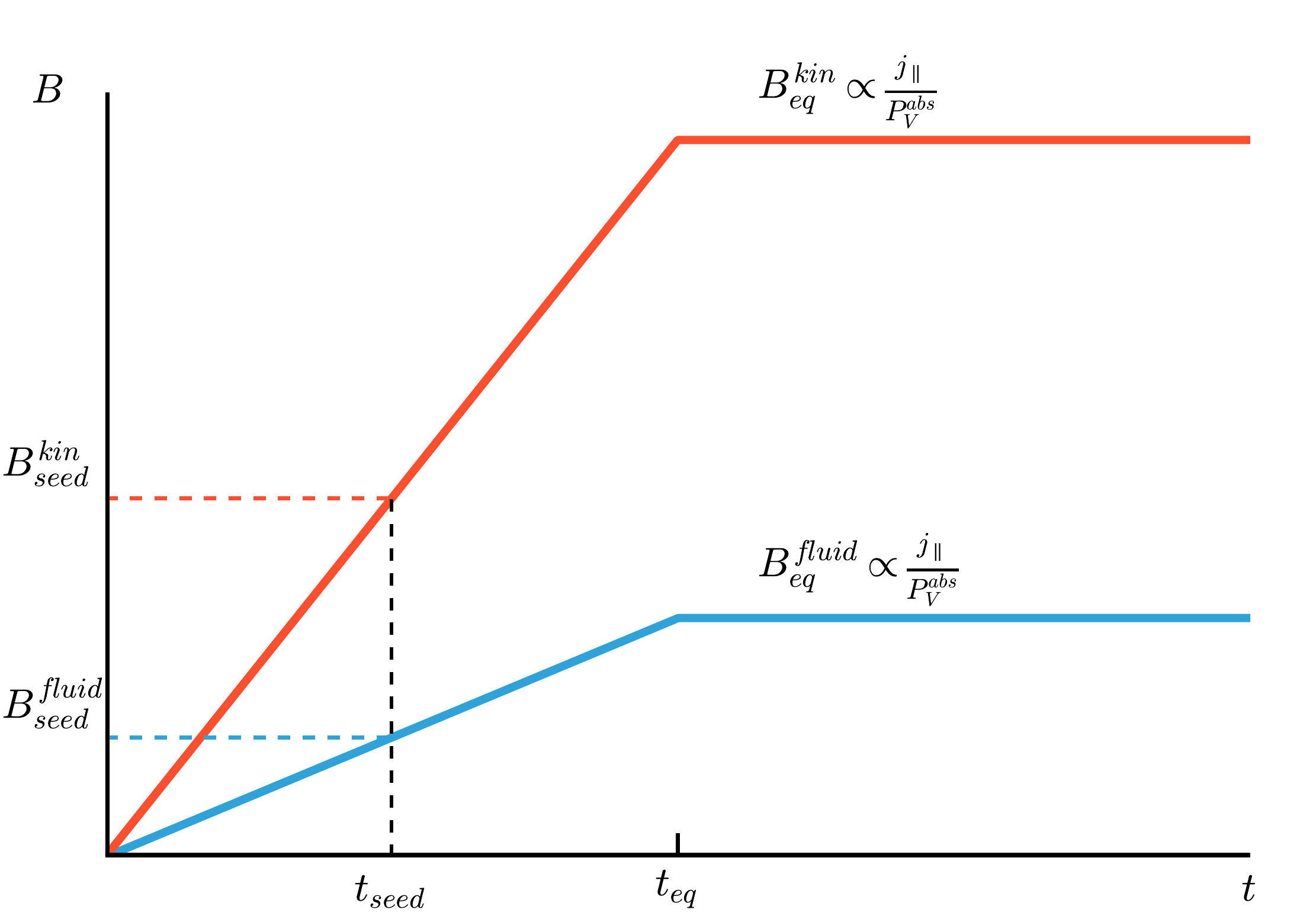}
\caption{\label{fig02}Schematic picture of the magnetic field growth. Magnetic field grows approximately linearly with time until it saturates at equilibrium value $B_{eq}$. Kinetic effects increase $B_{eq}$ to the same extent as they increase the current drive efficiency, but they hardly change the time to reach equilibrium $t_{eq}$. The actual value of the magnetic seed is determined at some characteristic time $t_{seed}\ll t_{eq}$, and it increases to the same extent as the efficiency increases.}
\end{figure}

\noindent Kinetic effects change the current drive efficiency and
thus the equilibrium field, but they do not affect much the time to reach equilibrium $t_{eq}$, which is the so-called "L/R time".   
That time is still determined
by the Spitzer conductivity, since the full distribution function is
equally pushed by an electric field \cite{Fisch1985}. The time to reach equilibrium $t_{eq}$ greatly exceeds the age of the universe, and so the actual value of the magnetic seed is determined at some characteristic time $t_{seed}\ll t_{eq}$, and
it increases to the same extent as the efficiency increases (see Fig.~\ref{fig02}).

\subsection{Blackbody incoming radiation and cyclotron absorption}

To take one example, let us assume that the incoming radiation from
the first slab is blackbody:

\begin{equation}
I_{\omega\Omega}=\frac{\omega^{2}}{8\pi^{3}c^{2}}\frac{T_{1}}{\bar{\gamma}\left(1+\bar{\beta}\cos\theta\right)}.
\end{equation}

\noindent If the plasma were already immersed in a parallel magnetic field,
then one of the absorption mechanisms would be synchrotron absorption.
In the non-relativistic case, it is determined by the emissivity \cite{Bekefi1966}:
\begin{equation}
\eta_{\omega\Omega}\left(\boldsymbol{\beta}\right)=\frac{e^{2}\beta_{\perp}^{2}\omega^{2}}{4\pi c}\left(1+\cos^{2}\theta\right)\delta\left[\omega_{c2}-\omega\left(1-\beta_{\Vert}\cos\theta\right)\right].\label{eq:cyclotron_emis}
\end{equation}

After some algebra it is easy to show that the current drive efficiency
in this case is
\begin{equation}
\frac{j_{\parallel}}{P_{V}^{abs}}=-\frac{\left\langle \beta_{\perp}^{2}\beta^{3}I_{2}\left(\beta_{\parallel}\right)\right\rangle +3\left\langle \beta_{\perp}^{2}\beta\beta_{\parallel}I_{1}\left(\beta_{\parallel}\right)\right\rangle }{6\left\langle \beta_{\perp}^{2}I_{1}\left(\beta_{\parallel}\right)\right\rangle },\label{eq:eff}
\end{equation}
where the averaging is over the initial distribution that
is taken to be a Maxwellian, and the following integrals are introduced:
\begin{align}
 & I_{1}\left(\beta_{\parallel}\right)=\int_{-1}^{1}dx\frac{1+x^{2}}{\left(1-\beta_{\Vert}x\right)^{3}\left(1+\bar{\beta}x\right)},\\
 & I_{2}\left(\beta_{\parallel}\right)=\int_{-1}^{1}dx\frac{x\left(1+x^{2}\right)}{\left(1-\beta_{\Vert}x\right)^{3}\left(1+\bar{\beta}x\right)}.
\end{align}

\noindent Keeping only the terms of the order $O(\beta_{\Vert}^{2}),$
$O\left(\bar{\beta}^{2}\right),$ $O\left(\beta_{\Vert}\bar{\beta}\right)$
we find:
\begin{equation}
\frac{j_{\parallel}}{P_{V}^{abs}}=\frac{\left\langle \beta_{\perp}^{2}\beta^{3}\right\rangle +9\left\langle \beta_{\perp}^{2}\beta\beta_{\parallel}^{2}\right\rangle }{15\left\langle \beta_{\perp}^{2}\right\rangle }\bar{\beta}\approx2.4\beta_{th}^{3}\bar{\beta}.\label{eq:eff_cyclotron_1}
\end{equation}
If we ignored the second term in the numerator of Eq.~(\ref{eq:eff_cyclotron_1})
and also did not account for $\beta_{\perp}^{2}$ in the absorption,
then the efficiency would be given by Eq.~(\ref{eq:eff_PR}), i.e.,
correspond to the case of Thomson scattering analyzed through fluid
theory.

From comparison of Eq.~(\ref{eq:eff_cyclotron_1}) and Eq.~(\ref{eq:eff_PR}),
we see that for cyclotron absorption the inclusion of the kinetic
effects boosts the generated current by a factor of 6. This is not
a huge change, though it is not insignificant either. For reference,
the fluid estimates for the galactic magnetic field are about $10^{-19}\:\textrm{G}$
\cite{Widrow2002}, while the estimates for the required lower bound
on the seed galactic field is about $10^{-14}\:\textrm{G}$ \cite{KulsrudZweibel2008}. 

Cyclotron absorption mechanism needs some parallel (toroidal) magnetic
field to be already present to work. However, we see that the efficiency
(\ref{eq:eff_cyclotron_1}) is independent of the magnetic field,
so it seems that we can get poloidal magnetic field from a very small
toroidal field, generated, for example, by the Biermann battery effect
\cite{Biermann1950}. This works only when all the incoming radiation
is absorbed within the plasma, so that the effective absorption length
is less than the characteristic size of the system. For blackbody
incoming radiation flux and cyclotron absorption the effective absorption
coefficient is
\begin{equation}
\alpha=\frac{4}{3\pi}\frac{k_{B}^{4}}{c^{3}\sigma_{SB}T_{1}^{3}}\omega_{p2}^{2}\omega_{c2}^{2},
\end{equation}
or $\alpha\approx10^{-20+n+2b-3k}\:\textrm{cm\ensuremath{^{-1}}}$
for $T_{1}/k_{B}=10^{k}\:\textrm{K}$, $n_{2}=10^{n}\:\textrm{cm\ensuremath{^{-3}}}$,
and $B_{2}=10^{b}\:\textrm{G}$. If we take typical protogalactic
values $T_{1}/k_{B}=10^{4}\:\textrm{K}$ and $n=1\:\textrm{cm\ensuremath{^{-3}}}$,
then for $B_{2}=10^{-20}\:\textrm{G}$ that realistically can be produced
by the Biermann battery the effective absorption length becomes $R\sim10^{71}\:\textrm{cm}$,
which is much larger than characteristic size of the system. Thus,
the cyclotron absorption mechanism cannot be responsible for the generation
of the galactic seed field. However, it might be a very effective mechanism
of  poloidal magnetic field generation in already highly magnetized
objects such as neutron stars.

\subsection{Cyclotron incoming radiation and absorption}

So far we considered that the incoming radiation is  blackbody. 
We can expect that if the incoming radiation were narrower
in $k_{\parallel}$, then its absorption would be more asymmetric
in parallel velocity of the second slab, which would result in 
enhanced efficiency.

To investigate this, consider the case where each of the slabs is immersed in an axial magnetic field, though the respective magnetic fields are not  not necessarily of the same strength. 
Suppose that cyclotron radiation is emitted by an optically thin surface layer of depth $L$. 
The incoming flux is then given by \cite{Bekefi1966}

\begin{equation}
I_{\omega\Omega}=\frac{n_{1}Le^{2}\beta_{th1}}{4\pi\sqrt{2\pi}c}\frac{\omega}{\left|\cos\theta\right|}\left(1+\cos^{2}\theta\right)e^{-\frac{\left(\frac{\omega-\omega_{c1}}{\omega\cos\theta}+\bar{\beta}\right)^{2}}{2\beta_{th1}^{2}}}.
\end{equation}

The current drive efficiency for cyclotron absorption has the same
form as Eq.~(\ref{eq:eff}), but with the following definition of
$I_{1}$, $I_{2}$:
\begin{align}
 & I_{1}\left(\beta_{\parallel}\right)=\left(\int_{-\infty}^{-\left|a\right|}dx+\int_{\left|a\right|}^{\infty}dx\right)\frac{1}{\left|x\right|^{3}}\frac{\left(x^{2}+a^{2}\right)^{2}}{\left(x-a\beta_{\Vert}\right)^{2}}e^{-\frac{\left(x+b\right)^{2}}{2}},\\
 & I_{2}\left(\beta_{\parallel}\right)=\left(\int_{-\infty}^{-\left|a\right|}dx+\int_{\left|a\right|}^{\infty}dx\right)\frac{a}{x^{4}}\frac{\left(x^{2}+a^{2}\right)^{2}}{\left(x-a\beta_{\Vert}\right)^{2}}e^{-\frac{\left(x+b\right)^{2}}{2}},
\end{align}
where $a=(\omega_{c2}-\omega_{c1})/\omega_{c2}\beta_{th1}\equiv\bigtriangleup\omega_{c}/\omega_{c2}\beta_{th1}$,
and $b=\left(\omega_{c1}/\omega_{c2}\right)(\beta_{\Vert}/\beta_{th1})+\bar{\beta}/\beta_{th1}$.

The first term in the denominator of Eq.~(\ref{eq:eff}) with $I_{1}$
and $I_{2}$ defined above is due to direct parallel momentum injection
and so depends on the sign of $\bigtriangleup\omega_{c}$, while the
second is due to asymmetric heating. Since now absorption is localized
in velocity space and most of the absorbed power goes into heating
rather than giving a parallel push, the second term completely dominates,
and the efficiency becomes essentially independent of the sign of
$\bigtriangleup\omega_{c}$. There are two qualitatively different
cases that produce current of different sign: $\left|a\right|\ll1$
(positive current) and $\left|a\right|\gtrsim1$ (negative current).
From numerical treatment it appears that for wide range of parameters
the efficiency is approximately given by 
\begin{equation}
\frac{\left|j_{\parallel}\right|}{P_{V}^{abs}}\sim10^{2}\beta_{th}^{2}\bar{\beta}.
\end{equation}

This is $10^{2}/\beta_{th}$ larger efficiency than that for the blackbody
radiation, for $T/k_{B}\approx10^{4}\:\textrm{K}$ is about $\sim10^{5}$.
Therefore, at least in the case when the plasma already possesses
some toroidal magnetic field, one can expect the generated poloidal
magnetic field to be orders of magnitude larger than the previous
estimates based on the fluid theory.

These results can be understood from the following qualitative picture.
The non-relativistic cyclotron resonance condition for an electron
of slab 2 moving with velocity $\beta_{\Vert2}$ to absorb the radiation
emitted by an electron of slab 1 moving with velocity $\beta_{\parallel1}$
is 
\begin{equation}
\omega_{c1}-k_{\Vert}c\left(\beta_{\Vert2}-\beta_{\Vert1}\right)=\omega_{c2}.\label{eq:res_condition}
\end{equation}

\noindent Here we use $k_{\parallel}$ corresponding to the reference
frame where the electron of slab 1 is stationary, and velocities $\beta_{\Vert1}$,
$\beta_{\Vert2}$ are measured in the frame where slab 2 is stationary.

\begin{figure}[]
\subfloat[\label{fig03_a}]{\includegraphics[width=0.24\textwidth]{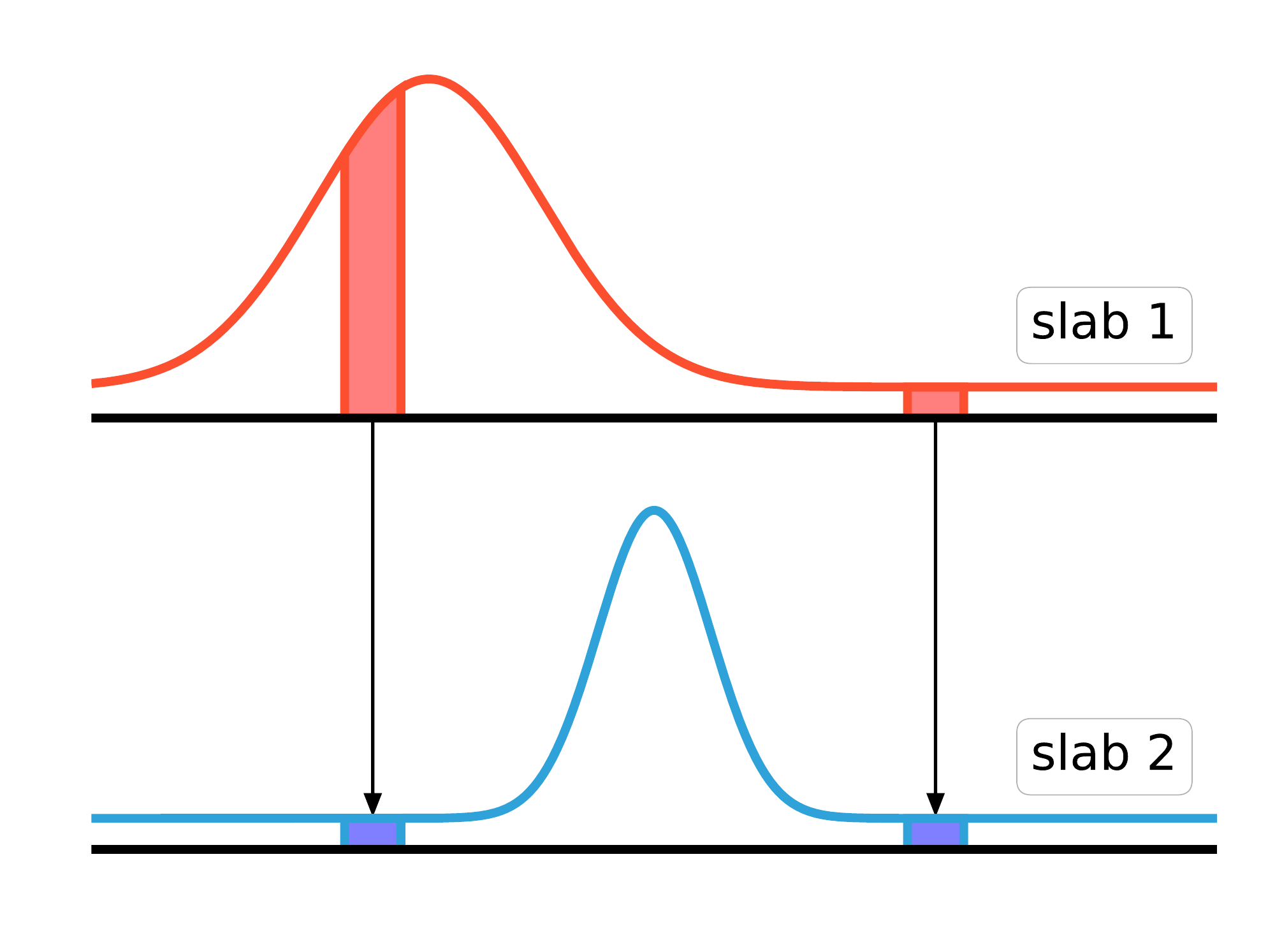}}
\subfloat[\label{fig03_b}]{\includegraphics[width=0.24\textwidth]{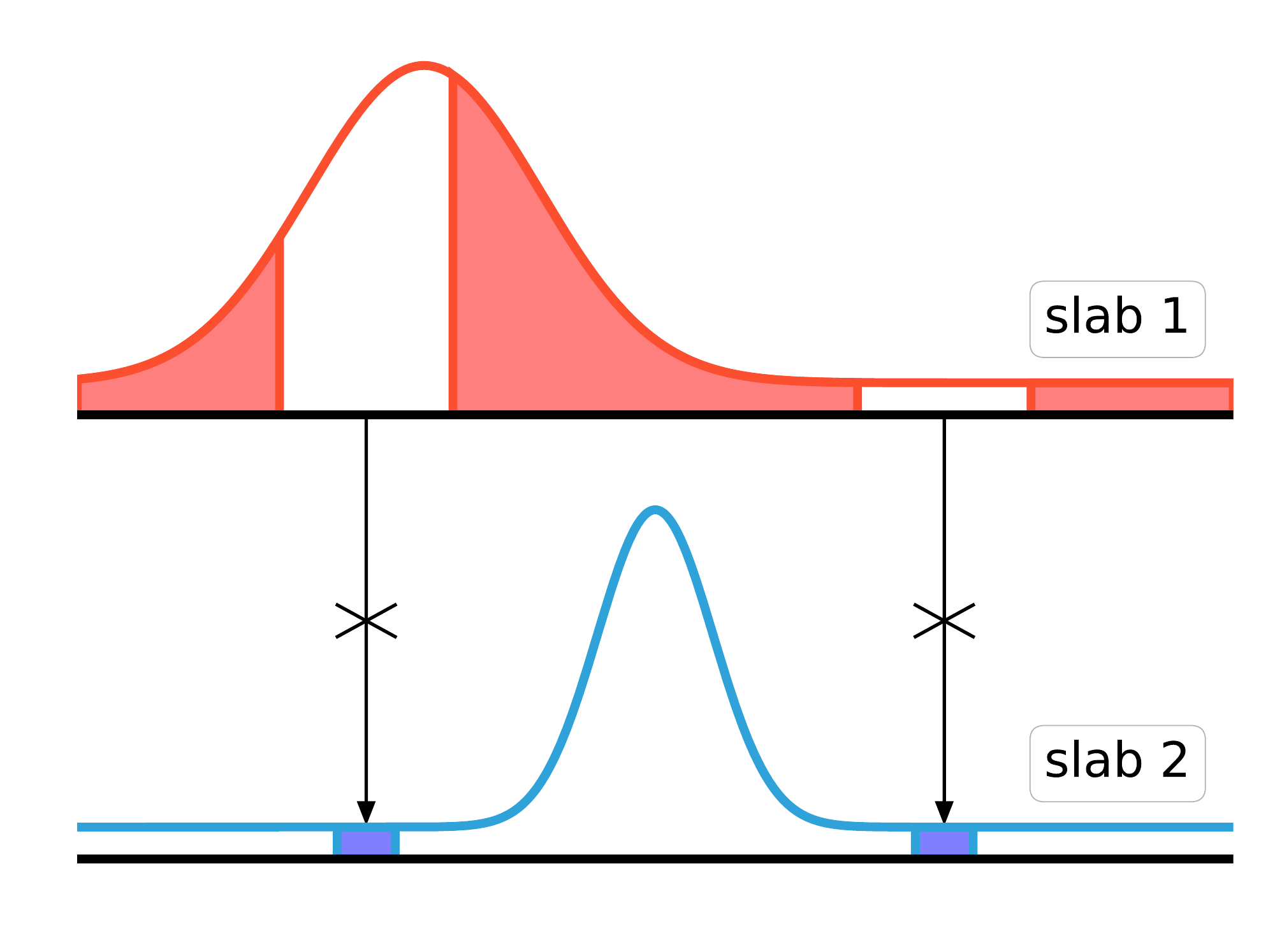}}
\caption{\label{fig03}(a) $\omega_{c1}\simeq\omega_{c2}$: electrons of slab
2 with negative velocities around $-\bar{\beta}$ (left blue region)
interact with the large number of electrons of slab 1 (left red region),
while symmetric electrons of slab 2 with positive velocities around
$\bar{\beta}$ (right blue region) interact with small number of electrons
of slab 1 (right red region). (b) $\omega_{c1}\protect\neq\omega_{c2}$:
electrons of slab 2 with negative velocities around $-\bar{\beta}$
(left blue region) has large number of electrons of slab 1 in the
non-absorption window (left white region) and thus absorb less energy
than symmetric electrons of slab 2 with positive velocities around
$\bar{\beta}$ (right blue region), which have small number of electrons
of slab 1 in the non-absorption window (right white region).}
\end{figure}

If $\left|a\right|\ll1$, then essentially $\omega_{c1}\simeq\omega_{c2}$
and the resonance condition is $k_{\Vert}=0$ or $\beta_{\Vert2}=\beta_{\Vert1}$.
The former condition does not depend on the velocities and cannot
lead to the asymmetry, the latter condition leads to an asymmetric
absorption. Indeed, the electrons with positive parallel velocity
$\beta_{\Vert2}\approx\bar{\beta}$ absorb less power than the electrons
with negative parallel velocity $\beta_{\Vert2}\approx-\bar{\beta}$,
because the latter are in resonance with a much larger number of electrons
in slab 1. Thus the electrons with negative parallel velocities will
experience less Coulomb drag force than the electrons with positive
velocities resulting in positive current. This situation is shown
in Fig.~\ref{fig03}\subref{fig03_a}.

If $\left|a\right|\gtrsim1$, then the magnetic fields are different
$\omega_{c1}\neq\omega_{c2}$ and the resonance condition (\ref{eq:res_condition})
implies that the electrons of slab 2 with velocity $\beta_{\Vert2}$
will resonantly interact with the electrons of slab 1 with parallel
velocities satisfying

\begin{equation}
\begin{cases}
\beta_{\Vert1}<\beta_{\Vert2}-\left|\triangle\omega_{c}\right|/\omega_{c1},\\
\beta_{\Vert1}>\beta_{\Vert2}+\left|\triangle\omega_{c}\right|/\omega_{c1}.
\end{cases}\label{eq:interaction_vals}
\end{equation}

\noindent Thus there is a window in the absorption for each electron.
The electrons of slab 2 with negative parallel velocity around $-\bar{\beta}$
will have larger number of electrons of slab 1 in this window and
thus will receive less power than the electrons of slab 2 around $\bar{\beta}$.
The result is negative current. This situation is shown in Fig.~\ref{fig03}\subref{fig03_b}.

Notice that one gets the same efficiency but with different sign for
the current driven in the first slab. For blackbody incoming radiation,
this current would be always in the opposite direction, but, interestingly,
for cyclotron radiation, it is possible to have the situation when
currents in both slabs flow in the same direction. Indeed, since the
parameter $a$ depends on temperature it can have different values
corresponding to two different regimes ($\left|a\right|\ll1$ and
$\left|a\right|\gtrsim1$) in each slab. Thus, we reach the surprising result 
that, for a differentially rotating plasma disk immersed in a toroidal
magnetic field and with a temperature gradient ,  it is possible that a toroidal
current will be self-consistently generated in the same direction throughout the disk.

\begin{figure}[]
\includegraphics[width=8.6cm]{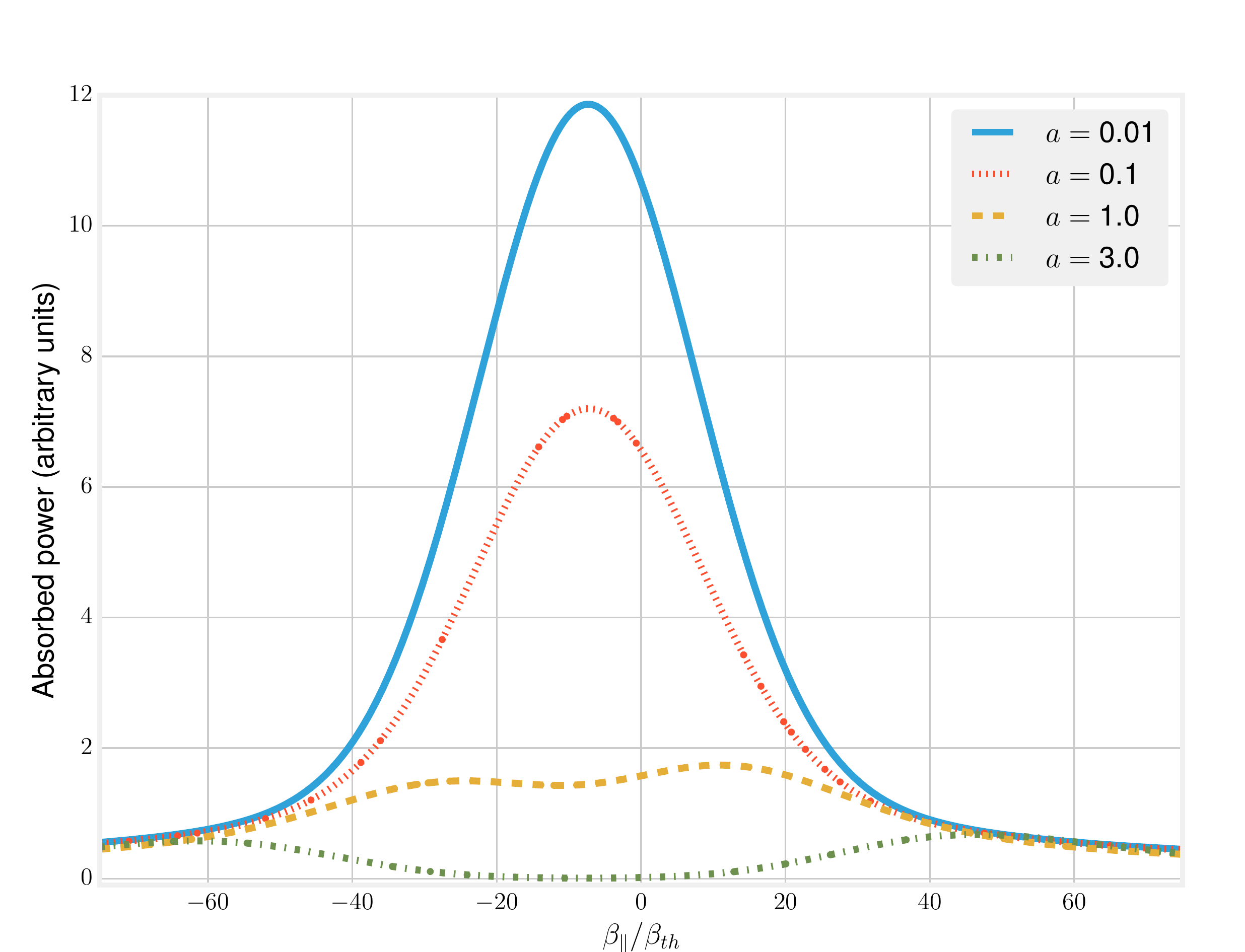}
\caption{\label{fig04}Absorbed power density per electron as a function of
parallel velocity for $\beta_{th}=0.01$, $\bar{\beta}=0.05$ and
different values of $a$: $a=0.01$ (solid blue), $a=0.1$ (dotted
red), $a=1.0$ (dashed orange), $a=3.0$ (dash-dotted green).}
\end{figure}

Figure \ref{fig04} shows the absorbed power per electron as a function
of the parallel velocity for $\beta_{th}=0.01$, $\bar{\beta}=0.05$
and four different values of parameter $a$. We can clearly see that
the absorption is asymmetric. For $\left|a\right|\ll1$ the situation
is basically analogous to the case of equal magnetic fields shown
in Fig.~\ref{fig03}\subref{fig03_a} when the electrons with negative
parallel velocities (around $-\bar{\beta}$) absorb more power resulting
in positive current. In contrast, for $\left|a\right|\gtrsim1$ there
is a dip in the absorption for the electrons moving with negative
velocities resulting in negative current (see Fig.~\ref{fig03}\subref{fig03_b}).
We can also see that, as the difference between magnetic fields of
the slabs increases, i.e.,\ as the parameter $\left|a\right|$ increases,
the total absorbed power decreases rapidly. Thus, in the limit $\left|a\right|\gg1$,
the efficiency should be viewed as questionable, because the total
absorbed power density is negligible and radiation has to pass through
a very large volume of plasma to be fully absorbed.

\section{Conclusion}

The generation of cosmic magnetic fields due to radiation transfer
can be significantly larger when one takes into account kinetic effects
rather than simply relying on fluid theory. In the case where the
radiation is from cyclotron radiation, namely, when there already exists
an ambient magnetic field, an increase in fields perpendicular to
the ambient field can be orders of magnitude larger when kinetic effects
are considered. Curiously, in the case of inhomogeneous field, it
is possible to generate these perpendicular fields such that the current
produced within two differentially traveling, radiating, and absorbing
slabs is in the same direction, an effect that would not be captured
in the fluid theory.

The formalism advanced here  shows how to  deal with a radiative process, which is kinetic by its very nature.
It is expected that the formalism advanced  here  can be applied to various areas of astrophysics where
radiative kinetic effects might be important, for example, to radiative magnetic
reconnection \cite{Uzdensky2016}. It is also hoped that the approach taken here  might help to make more accurate  the currently widely used astrophysical radiative transfer codes, which, to the best of our knowledge, exist only 
in the hydrodynamic version (see, for example, Refs. \cite{ShaneStoneJiang2012,ZEUSIII}).

\section*{Acknowledgement}

This work is supported by DOE Contract No. ~DE-AC02-09CH1-1466.

\bibliographystyle{apsrev4-1}
\bibliography{Radiative_transfer_dynamo_effect}

\end{document}